# "They're all about pushing the products and shiny things rather than fundamental security":
# Mapping Socio-technical Challenges in Securing the Smart Home

Jiahong Chen[1] and Lachlan Urquhart[2]

## Abstract

Insecure connected devices can cause serious threats not just to smart home owners, but also the underlying infrastructural network as well. There has been increasing academic and regulatory interest in addressing cybersecurity risks from both the standpoint of Internet of Things (IoT) vendors and that of end-users. In addition to the current data protection and network security legal frameworks, for example, the UK government has initiated the "Secure by Design" campaign. While there has been work on how organisations and individuals manage their own cybersecurity risks, it remains unclear to what extent IoT vendors are supporting end-users to perform day-to-day management of such risks in a usable way, and what is stopping the vendors from improving such support. We interviewed 13 experts in the field of IoT and identified three main categories of barriers to making IoT products usably secure: technical, legal and organisational. In this paper we further discuss the policymaking implications of these findings and make some recommendations.

## Introduction

The security of smart home products has been an increasing concern for consumers and policymakers alike. As the number of smart devices connected to the Internet of Things (IoT) is growing steadily, more cybersecurity incidents associated with smart devices are being reported. In theory vendors of IoT products would have incentives to ensure their products are safe to use 'out of the box' and also provide help for consumers to manage their smart home security across the life cycle of the technologies. In reality, however, this is not taking place at a satisfactory level, and the resulting legislative shifts are changing the regulatory landscape in the UK and EU. Our investigation unpacks the socio-technical causes of poor smart device cybersecurity, criticises the current regulatory frameworks and explores alternative approaches to improving the state of this domain. We interviewed 13 experts from different sectors all of whom had significant experience in the area of IoT security. In this article, we will first provide some background information about the regulatory landscape in this field as well as an explanation as to why this inquiry is important at this time, as governance of smart homes and cybersecurity is changing. The design and results of our study will then be presented, before we further discuss how such insights can help us make sense of the technical, legal and organisational barriers that the industry is facing, and also the regulatory support that might be needed to tackle these barriers.

[1] Lecturer in Law, University of Sheffield, UK. Email: jiahong.chen@sheffield.ac.uk
[2] Lecturer in Technology Law, University of Edinburgh, UK. Email: lachlan.urquhart@ed.ac.uk

## Context and motivation

### *The popularity of smart home products and the rise of associated cybersecurity threats*

Consumer IoT is a fast-growing industry both in the UK and worldwide. The global smart home market was valued at $43.4bn in 2017 and estimated more than double at $91bn for 2020 (Basul, 2020). In the UK, the smart device market has seen a strong growth as well. A 2020 report showed that the market penetration of smart speakers had grown from 7% in 2017 to 29% to 2020, and smart TVs also saw a growth from 39% to 49% for the same period (TechUK, 2020).

The growth of spending on smart devices in 2020 has been affected by the COVID-19 pandemic, due to economic uncertainties, restricted retail opportunities, and disrupted manufacturing and distribution (ABI Research, 2020). In the long term, however, it is believed that the pandemic will further accelerate the adoption of smart home products for various reasons, including dramatic increase in time spent at home and the higher demands in tracking health and fitness (Deloitte, 2021).

Concurrently, the sheer increased number of connected devices raises concerns about the heightened cybersecurity risks. Vulnerable smart devices without security provisions may become targets of cyberattacks, which pose threats not just to the end-user's security and privacy (Ward, 2014; BBC, 2020b; BBC, 2020a), but also to the operation of the infrastructural network. There is a concern around the scale of cybersecurity attacks. Whilst local smart home cybersecurity risks can directly impact physical or informational security of home occupants, there is scope for these risks to scale up. For example, if a zero-day vulnerability exists across IoT devices installed in thousands of homes, there could be wide-scale disruption and danger. Since the Mirai DDoS attack began in 2016, there have been continual waves of botnet attacks exploiting vulnerable devices on the IoT. In 2020 alone, security researchers have identified multiple botnet variants, such as Hoaxcalls (Arghire, 2020), Dark_nexus (Osborne, 2020) and Mukashi (Lakshmanan, 2020). If a large number of IoT devices become compromised, they can be weaponised against critical facilities, including healthcare services (Miliard, 2016) and power grids (Malan et al., 2020). Threats to smart home devices can take a variety of forms. Barnard-Wills et al. (2014) have identified 9 main categories of threats,[3] while (Malan et al., 2020) have identified 13.[4] A recent report finds that IoT-based DDoS attacks have been intensified during the current pandemic, as more devices have been connected to the network and organisations have struggled to maintain the workforce (Hope, 2021).

---

[3] The 9 categories include "Legal", "Nefarious Activity/Abuse", "Eavesdropping/Interception/Hijacking", "Outages", "Physical attacks", "Unintentional damages (accidental)", "Disasters" and "Damage/Loss (IT Assets)".

[4] The 13 categories include "Physical attacks", "Distributed Denial of Service (DDoS)", "Unintentional damage (accidental)", "Failure/malfunctions", "Outages", "Eavesdropping/interception/hijacking", "nefarious activity", "Spoofing", "Tampering", "Repudiation of actions", "Information disclosure", "Elevation of privilege" and "Unsupported endpoint management".

*The "Secure by Design" initiative and the broader regulatory context*

In the light of the ongoing cybersecurity threats associated with vulnerable consumer IoT products, the UK government published the Secure by Design report in 2018, which calls for "a fundamental shift in approach to moving the burden away from consumers having to secure their internet connected devices and instead ensure strong cyber security is built into consumer IoT products and associated services by design" (DCMS, 2018b). The report proposed a Code of Practice for Security in Consumer IoT Products and Associated Services, which was adopted later that year with a shorter title of "Code of Practice for Consumer IoT Security" (DCMS, 2018a). The Code contains 13 guidelines, which include:

- "no default passwords",
- "implement a vulnerability disclosure policy",
- "keep software updated",
- "securely store credentials and security sensitive data",
- "communicate securely",
- "minimise exposed attack surfaces",
- "ensure software integrity",
- "ensure that personal data is protected",
- "make systems resilient to outages",
- "monitor system telemetry data",
- "make it easy for consumers to delete personal data",
- "make installation and maintenance of devices easy" and
- "validate input data".

It should however be noted that all these guidelines are currently provided as "good practice" but not mandatory. Whether and how they should be mandated in the legislation is separate issue that we will discuss below.

In order to set out the statutory, baseline security requirements for consumer IoT products, the government launched a public consultation on its regulatory proposals in 2019 (DCMS, 2019). The consultation set out three main regulatory options, which can be summarised as:

- "mandating a security label" (the government's preferred option),
- "mandating the top 3 guidelines" (i.e. "no default passwords", "implement a vulnerability disclosure policy" and "keep software updated"), and
- "mandating all 13 guidelines".

After the consultation, an updated version of the regulatory proposals was published in 2020, which essentially shifted the preferred regulatory option to the second one, i.e. mandating the top 3 guidelines set out in the Code of Practice as the baseline requirements (DCMS, 2020c). In April 2021, the government responded to the consultation (DCMS, 2021), confirming they will proceed with UK wide legislation and establish an enforcement body that will ensure protection for consumers from insecure connected consumer products including smart speakers, televisions, doorbells and phones (but not desktops or laptop computers).

Furthermore, whilst the UK Cybersecurity Strategy continues beyond 2021, it remains unclear when the current Strategy ends. A progress report from 2020 highlights that making technology secure by design, particularly for consumer IoT, has been a key priority. They state "A core tenet of the regulatory approach is to implement transparency between those who make, stock and sell IoT devices" and they will push for legislation by the end of the Strategy, alongside supporting international standard development, like the success of ETSI EN 303 645 (Cabinet Office, 2020: 21). Parallel developments are taking place in the EU, where there have been discussions on regulating connected devices either through a new legislative framework (ZVEI, 2018; DIGITALEUROPE, 2019) or by reforming the Radio Equipment Directive (BEUC, 2019), but no formal proposal has been tabled yet. The recent proposal on regulating AI systems (Commission, 2021a) may also be relevant given that AI solutions are increasingly applied to IoT systems, and that the draft has a relatively broad definition of "operators" of AI systems.

The government's regulatory initiative should be viewed against the backdrop of the broader cybersecurity and data protection legislation, as well as the associated standardisation efforts. In relation to cybersecurity law, the EU's Directive on Security of Network and Information Systems (NIS Directive) was adopted and entered into force in 2016, which was implemented in the UK through the Network and Information Systems Regulations 2018 (NIS Regulations). The NIS legal framework sets out the criteria for the identification of operators of essential services, covering the energy, transport, banking, financial market infrastructures, health, drinking water and digital infrastructures sectors, as well as the key digital services, i.e. online marketplace, online search engine and cloud computing services. It also lays down the minimum security and incident notification requirements, and the role and cooperative mechanisms of national competent authorities and incident response teams. Manufactures, retailers and service providers of consumer IoT products are not typically subject to the NIS framework, but if an attack causes disruptions to an essential sector or a key digital service, the contingency measures set out in the legislation should be put in place. It is a different question whether the IoT sector *ought to be* covered by the NIS framework (or at least the key players in the sector), given the serious impact of connected devices on the security of the wider network, a topic we will discuss later.

In terms of data protection law, the EU's General Data Protection Regulation (GDPR, 2016) may come into play to the extent that personal data is involved in the use of consumer IoT devices. Article 5(1)(f) speaks of the principle of "integrity and confidentiality", requiring "appropriate security of the personal data, including protection against unauthorised or unlawful processing and against accidental loss, destruction or damage, using appropriate technical or organisational measures". In addition, Article 25(1) imposes a "data protection by design" requirement on data controllers, who are required to "both at the time of the determination of the means for processing and at the time of the processing itself, implement appropriate technical and organisational measures […] in an effective manner and to integrate the necessary safeguards into the processing in order to meet the requirements of this Regulation and protect the rights of data subjects." The combination of technical and organisational measures is also mandated as part of the "security of processing" duty under Article 32 in order "to ensure a level of security appropriate to the risk". Yet, despite the shared "by design" phraseology between the GDPR and the Secure by Design initiative, there seems to be a misalignment between the two fields when it comes to the actual meaning of "by design". The three GDPR provisions cited above have all clearly envisaged technical and organisations measures as effective ways to comply with data protection principles. As regards the UK government's Secure by Design approach, however, it is defined as "[a] design-stage focus on ensuring that security is in-built within consumer IoT products and connected services" (DCMS, 2018b: 33). This overwhelming focus on the *technical* design is also evident in the Code of Practice and the subsequent regulatory proposals. It is unclear what the organisational dimensions are in the development of consumer IoT products, how much these aspects matter in managing cybersecurity risks, and whether the disregard of these aspects will warrant regulatory intervention. All these questions form an important part of what motivates the inquiry of this paper.

### *The role of IoT vendors and the barriers they are facing*

There is a rich body of literature on cybersecurity management within commercial organisations. Plenty of research has been devoted to combating cybercrimes on the technical front, including the common threats organisations are facing (Saleem et al., 2017; Tounsi and Rais, 2018; Spremić and Šimunic, 2018), the technological challenges in addressing those threats (Stanciu and Tinca, 2017; Dambra et al., 2020; Zlomislić et al., 2017) and strategies to mitigate such challenges (Clim, 2019; Tselios et al., 2020; Chan et al., 2019). More recently, research has also focused on the human factors in managing cybersecurity risks for organisations of various sizes. Oltramari et al. (2016), for example, underlines the importance of addressing cybersecurity as a socio-technical system: "untangling the complexity of cyber security does not solely depend on pinning down the computational elements into play, but demands a thorough analysis of the human factors involved." (See also: Vieane et al. (2016); Oltramari et al. (2016)) They have identified the defender, user and attacker as the main actors and mapped out a range of internal and external characteristics affecting cybersecurity risk assessment. Van Zadelhoff (2016) went even as far as to comment that "the biggest cybersecurity threats are inside your company". Work by Bowen et al. (2011) focuses on larger organisations, whereas Bada and Nurse (2019)'s work puts forward strategies intended for SMEs. Similarly, Norval et al. (2019) found that start-ups were particularly fearful about fines and liabilities arising from security vulnerabilities and that whilst cyber-attacks are a major concern for IoT firms, that use of major platforms gave start-ups reassurance e.g. for use of their authentication systems. Across the board, Sirur et al. (2018) identified disparities among organisations with varying sizes and access to resources in their reactions to compliance with the GDPR. A 2020 review by the UK government has surveyed the common barriers faced by organisations managing cybersecurity risks, identifying notably the inabilities preventing them from taking actions, the lack of commercial rationale to invest in security, and the complex and insecure digital environment (DCMS, 2020a).

In the context of domestic IoT, there are discussions on the role of human actors, who might turn out as the source of cybersecurity threats or vulnerabilities as much as who might fall victims to such threats. As Leukfeldt and Yar (2016) argue , routine activity theory (RAT) (Cohen and Felson, 1979) can be a valuable tool in managing cybercrime by understanding the nature of motivated offenders, suitability of targets and capability of guardians to mitigate cybercrimes. Piasecki et al. (2019) draw on this analytical framework to understand how to manage human and technical threats associated with smart home technologies, examining not just threats from outside the home but also from within. From a risk management point of view, human factors in smart home can manifest around issues such as usability (Parkin et al., 2019), awareness (Park et al., 2019), skills (Zhang et al., 2018), resources (Lin and Bergmann, 2016) and motivation (Zaidan and Zaidan, 2020). These factors all affect the effectiveness of the users' efforts in self-managing security risks in their smart homes.

To sum up, the literature addresses the socio-technical aspects of cybersecurity management mainly from two aspects: How businesses manage threats to *their own organisations*, and how end-users manage threats to *their own homes*. What seems missing is the socio-technical dynamics involved in *businesses supporting end-users* to manage threats. This is especially critical in the context of domestic IoT because the control over the functioning of smart devices is distributed across commercial and domestic actors. From a data protection perspective, for example, Chen et al. (2020) have discussed the legal complexities involved in complying with data protection law in a smart home setting. The development and operation of domestic IoT products often involve a multitude of players, including architectural developers, third-party component builders, device manufacturers and end-users. They all exercise some extent of control over how the system functions, but in starkly different ways. The legal consequence of this phenomenon, in the light of the European Court of Justice's jurisprudence, is that end-users and IoT vendors (including manufacturers and service providers) are likely to be held jointly responsible for the processing of personal data by smart devices. Yet, it is unclear how the responsibilities should be allocated in practice to duly reflect their respective control, calling into question whether and how domestic data controllers should be treated differently from commercial controllers, given their weaker position in the market and their lack of necessary skills and resources to manage cybersecurity effectively (Urquhart and Chen, 2020; Janssen et al., 2020). This also raises the question as to how IoT businesses can support end-users to fulfil their data protection duties and to prevent cybersecurity threats, which may arguably form part of those businesses' accountability duty.

In a perfect market where transaction costs, information asymmetries and externalities are negligible, the involved parties would in theory come to an agreement whereby the best-positioned party – to whom the costs would be minimum – would take actions to minimise cybersecurity risks and get compensated by other parties. In reality, however, we are seeing a large amount of insecure, unsafe IoT products being sold to consumers. The question then is, **what is stopping organisations from developing products that come with usable security**? As discussed above, there is a gap in the literature on the socio-technical factors affecting an organisation's ability to support end-users to manage cybersecurity. In this study, we focus on the technical, legal and organisational barriers to a bigger role played by IoT vendors in securing the smart home products they offer.

## Methodology

In our study, semi-structured interviews have been carried out, by one single interviewer, during a period between August and December 2019 with 13 participants. These were dominantly with experts from backgrounds within the fields of the internet of things and cybersecurity from across civil society, government, industry and legal sectors. The first wave of interviewees were recruited through gatekeepers and our professional network of contacts, followed by a second wave through snowball sampling. In terms of background for the different organisations interviewed, this included civil society groups promoting responsible technology development; governmental agencies working on cybersecurity; legal services specialising in technology law. All participants had worked in or with at least one organisation developing IoT products or providing IoT services. For ethical considerations, the names of the participants or their organisations are not disclosed. Instead, they will are quoted with a list of pseudonyms below, combined with some non-identifying background information. To further assist with anonymity, the presumed gender of the pseudonym does not necessarily correspond to the gender identity of the participant. Key themes in our interview questions included the interface between regulation and design for smart home cybersecurity; definitional issues of what constitutes a smart home and the main security risks in the industry;

compliance strategies and structures within organisations to support secure smart device development; priority areas of law, standards and regulatory frameworks; best practice around regulating devices once in the marketplace; and the role of technical vs legal solutions to dealing with smart cybersecurity vulnerabilities. All interviews were audio-recorded (between 25 and 60 minutes), with the names and other direct identifiers of the participants removed from the recording before transcription. Coding was carried out with an inductive method, with one researcher first coding 7 transcripts and the other researcher 6 transcripts. The two researchers then discussed their code books and integrate them into one. With the new code book, both researchers re-coded the transcripts they initially coded, and then review each other's coding, adding additional annotations. This coding method was intended to avoid the dominance of the interpretation by one researcher and to allow collaborative coverage of all transcripts. By the time of publication of this article, all recordings have been permanently deleted in line with the study's ethical approval.

Table of pseudonyms and background information (Wikipedia, 2021)

| # | Pseudonym | Organisation type | Years of Experience | Role | Domain of expertise |
|---|---|---|---|---|---|
| 1 | Alice | Civil society | 4 | Senior policy advisor | IoT security |
| 2 | Bob | Civil society | 25 | Technology programme manager | Security and privacy |
| 3 | Charlie | Civil society | 25 | Legal counsel | Corporate service |
| 4 | David | Government | Unspecified | Researcher | Cybersecurity |
| 5 | Erin | Government | 4 | Researcher | Cybersecurity, software development |
| 6 | Frank | Government | Unspecified | Researcher | IoT security |
| 7 | Grace | Industry | 23 | Cybersecurity professional | Cybersecurity |
| 8 | Heidi | Industry | 40 | CTO | Communications, security |
| 9 | Ivan | Civil society | 10 | Legal counsel | Corporate service |
| 10 | Judy | Industry | 3 | Technical lead | IoT |
| 11 | Mallory | Industry | Unspecified | CTO | Digital services |
| 12 | Mike | Legal service | Unspecified | Managing director | Corporate service |
| 13 | Niaj | Civil society | 9 | Policy director | Privacy, surveillance |

## Findings

As mentioned above, our study focuses on the *technical*, *legal* and *organisational* dimensions of businesses providing IoT products with regard to how these factors affect the way they make their products secure and support users to improve security. This taxonomy finds support in recent work by Veale and Brown (2020), who have reviewed the interdisciplinary literature on cybersecurity, and have identified three main areas of cybersecurity research: "technical aspects", "human factors and social sciences" and "legal dimensions". In what follows, we present the findings from the interviews around these three themes, each addressing the threat landscape, barriers and solutions regarding that aspect.

### *Technical aspects*

Interviewees pointed out different types of threats insecure IoT devices might pose. For consumers, the threats identified concern the use of data and the safety of users. As government researcher Frank noted, *"there always have been risks if they had a computer in their house, but now the threat profile and the footprint of the house now is going to get larger."* Privacy was a commonly shared concern among interviewees but the angles are slightly different. Civil society legal counsel Charlie gave the example of unauthorised observation via compromised cameras (and so did legal counsel Ivan and law firm managing director Mike): *"If they could break into the security of my wireless network, they could turn on a camera and observe things in my house that I wouldn't want observed."* At the same time, he also raise the point about over-collection of personal data and the monetisation thereof, something shared by technology programme manager Bob and CTO Mallory, with different examples. Bob commented: *"That [IoT] data can be very invasive. And it's often combined with data from other companies or other devices to form a very explicit picture of the user that can then be sold."*

Potential harms to the physical safety of the users were also mentioned by interviewees. Ivan, technical lead Judy and Mike all gave their hypothetical examples how smart devices can be weaponised to inflict harms to the users.

*"For example, if you compromise someone's heating, that will - on its own it doesn't sound like a big deal because you could just turn it down but then there's the elderly, which may not be able to turn down their heating or turn up their heating. So, in winter, or the summer, that could actually be potentially lethal."* (Judy)

Identity theft was another threat picked up by legal counsels Charlie and Ivan. This is related to another point regarding breaches of confidentiality, which could be an even bigger issue if the compromised device has been connected to a workplace network from one's home (senior policy advisor Alice and Ivan).

From a wider, infrastructural perspective, many interviewees (Alice, Frank, Ivan and Mallory) shared the concern about vulnerable device being exploited for carrying out large-scale cyberattacks, with the Mirai botnet expressly mentioned by some interviewees. Frank noted that:

*"I think from the UK point of view, if we're talking about the nation, what our main concern is, is that a class of devices is affected which has a massive scaling impact on the UK. So if one person's machine or one person's device is taken down, yes, that one person is very annoyed, but we can deal with that. If somehow somebody affects all of one manufacturer's devices and they're popular in the UK that could have a massive impact on the UK economy and also UK trust and stuff like that."*

As regards what is causing these vulnerabilities, bad practices in the technical design of the product were considered a major barrier to securing smart home products:

*"I think, inherently in some cases, it's things like the lack of interface or the low powered devices, but also because of bad practices in things like companies setting default passwords, not providing security for payments and things like that."* (Policy director Niaj)

Another major technical challenge is how to avoid giving users warnings too often (Bob), some of which could even be false positives (CTO Heidi). The desensitisation of security alerts calls for further research into mechanisms to keep users informed but not overwhelmed.

In terms of available technical solutions to some of the challenges above, Alice and cybersecurity professional Grace explicitly mentioned Manufacturer Usage Description (MUD) as a promising approach. MUD is the formal specifications of the intended purpose of an IoT device, which enables its network behaviour to be monitored and restricted (Hamza et al., 2018). Developed as a standard to be implemented on terminal devices (IETF, 2019), which form the "edge" of the IoT, MUD also signifies an important part of the wider developments in edge-computing technologies, aiming to shift the computing activities closer to the edge, such as data processing and security management. Mallory pointed out edge computing could also be a helpful paradigm by minimising the risks associated with data transfers to the cloud:

*"So the classic example is if you put a camera into someone's home and you then stream that raw data in whatever format you've got a massive security problem. If you start to incorporate some edge computing into that and then you can say, what does the camera need to know? All I need to know is whether there is someone in this room or not or whether it's just a cat moving around, yeah?"*

The promise of edge-computing will be further discussed below in the "discussion and recommendations" section.

### *Legal aspects*

Apart from technical barriers, interviewees also reported major legal reasons why IoT vendors were not making all necessary efforts to ensure their products were secure. One common theme picked up by most interviewees was the fact that competing standards existed, making compliance with some or all of these standards more difficult. Such divergence in legal or normative standards could be a harmonisation issue, or a jurisdictional one.

The harmonisation issue arises when the guidance provided by different standard-setting bodies are inconsistent. Government researcher Erin summarised the issue as follows (which was similarly also mentioned by Alice):

*"But there's been an absolute mountain of competing sets of guidance really. Which means it's very difficult for somebody implementing something to know which to pick. If you're a start-up, you don't want to have that long lag where you're just trying to understand the regulatory environment and which guidance to follow, it can mean the difference between your company succeeding and failing."*

Perhaps the even more challenging issue is the jurisdictional one, where different countries may have their own standards, and vendors selling products to more than one country would have to decide whether to apply the strictest rules to their products sold to all the markets they trade, or to tailor the products to country-specific regimes. Bob captures the essence of the issue at hand as:

*"The certification programmes that they develop will not be mandatory for any of the EU member states. They will be an option. Anyone in an EU member state can choose to do anything or nothing with it. They can look at it and shrug their shoulders and say, not interested, move along. Or they can go all the way to the other extreme and make it mandatory to do business within a particular country."*

Charlie and Frank also reported similar observations.

Interviewees have also pointed out other limitations of the current regulatory framework as being one-size-fits-all, lagging behind reality, lacking baseline requirements and reliant on private enforcement:

*"And just because it's a small IoT device doesn't make it anything special, you'll need to comply with the rules."* (Frank)

*"It's just been a bit slow because these things are already in people's houses, but legislation is always lax, it's behind technology."* (Ivan)

*"My concern is that there isn't a regulatory framework in place to require compliance with many of these things because why does your protocol need to be secure? You can still sell your product. No-one's going to know about it for, you know, a year, two years, and then you realise it's being used in a botnet or something and then you try and fix it later on."* (Ivan)

*"I also don't think it's realistic to expect consumers to reinforce security law, or any other law, because let's face it, most people hate lawyers."* (Mike)

In the light of these legal barriers, interviewees made various suggestions on how these challenges can be at least partly addressed. Laying down uniformed, practical guidelines or standards seemed to be the most popular idea, shared by Alice, Grace, Ivan and Niaj. Despite the difference in the exact preferred terms, all those four interviewees seemed to be supportive of something that can be summarised by Alice as:

*"So we work with a coalition of governments that is working really hard to come up with best practices and kind of – I don't want to say standards, but I guess frameworks and baselines for security."*

Further regulation was also called for, including suggestions of a labelling or certification scheme (Alice, Grace and Ivan) or introducing a legally binding code of practice as proposed by the government (Frank). However, respondents showed divergent attitudes towards the effectiveness of engaging consumers by providing additional, accessible information. When discussing labelling schemes as a possible solution, for example, Grace expressed a degree of optimism while Mike was more sceptical (but interestingly, both picking up on the analogy with food safety regulation):

*"So I don't think we just don't provide consumers with enough information at the present moment in time, if you think about it, you know, to bring a food analogy which is often used in this space, we know what's in our food packaging, we know what the breakdown of calories is, we know all these pieces from the external of the box but when it comes to IT equipment, white box."* (Grace)

*"[Asked about the role of consumers:] No more so than I think it's appropriate for consumers to buy food from a supermarket and be responsible for checking it's fit for consumption. You talk to your average consumer, they will not have a clue about security of internet connected devices. We've just about got to a point where consumers know to look for a padlock in the URL bar of a browser. They don't know what it means."* (Mike)

In the discussion section below, we will further elaborate the possibility of expanding the scope of cybersecurity laws to cover smart devices, before exploring the arguably more market-friendly labelling and certification mechanisms.

*Organisational aspects*

To unpack the internal dynamics within an IoT vendor and their impact on the security of the consumer IoT products, the interviewees were asked questions about decision-making and product development across various departments within a typical organisation. A number of such organisational factors seem to have negatively impacted the security strategisation regarding their products.

First, misalignment of departmental priorities seems to be a major concern. Although, as Grace noted, in theory security should be a shared object across the entire organisation if its value for the organisation was properly understood, many interviewees reported that, in reality, the goals of the design/development department can come into conflict with those of the compliance/legal department. This is not helped by potential miscommunications between departments:

*"Yeah, my principle observation would be in many cases they don't speak the same language. And they have a very hard time communicating with each other. People often are not translating geek to English, or between technical and business environments. I think that's a real problem. People that are trained in the compliance and legal sphere don't – some of them may have a very deep technical understanding background, many of them don't. And likewise people that are very deep in the technology space often have a limited understanding of the legal and compliance issues."* (Bob)

*"I really think that there is a lack of communication a lot of time between technologists and business people. Not just for IoT but for everything. Where it can be really difficult for an engineer to go to someone who has no technical background and say, you are not clearing these risks that may not appear for five or ten years, and who are in a business position to accept those risks as something so urgent that they have to delay a product or delay a profit on their – or a return on their investment even."* (Alice)

Similar comments were made by Ivan, Judy and Mike. Such misalignment and miscommunications seem to have led to another related barrier: A corporate culture that treats security compliance very much as an afterthought, as highlighted by both Bob and Mike.

*"I think too often companies design products and services and view privacy and security and <u>compliance issues as something they can do as an afterthought at the end of a process</u>, it clearly doesn't – it needs to be part of a very – from the very start of the design process, all the way through, while the product or service is in use in the field, all the way through to end of life. Because it never works well if you view it as a bolt on that you can throw on at the end."* (Bob)

*"What you sometimes find is that <u>cybersecurity and privacy are a bit of an afterthought</u>, which is we've built this product, now can you tell me what I need to do to comply with data protection law, and then you've really lost the opportunity of saying, well go right back to the beginning, could you build it in a better way."* (Mike)

If short-term profitability becomes the top priority of an organisation, the objective of developing and manufacturing secure products might come across as an unjustified cost, especially considering the costs of longitudinal support. Mallory noted that:

*"I don't think that the original manufacturing costs of the product are the problem, I think it's the support that's given from that point onwards isn't it? I mean I think in a changing security*

*environment, if a business, the cost of business is going to be supporting the lifetime of that product rather than its manufacturing, so I don't think the cheapest of products in the first instance has really got that much to do with it."*

Ivan and Niaj had similar observations specifically in relation to SMEs. Many of them simply do not have the luxury to plan for long-term security management for their products, nor do they have specialised teams to offer legal advice. Driven by the need to survive as a start-up, many smaller IoT vendors have gone for the higher-risk decisions.

In this regard, perhaps a culture change would be what is most needed to heighten the overall level of smart product security for the entire economy. This was suggested by Grace:

*"First and foremost I think that the nature of that culture comes from the realisation that security can be an inhibitor, and if you realise that you can build security in it makes security no longer an inhibitor, so therefore it becomes an accelerator or an enabler."*

In terms of managerial structure and product development cycle, suggestions were made about deeper cross-department integration early on in the design process. Heidi suggested that product designs should be subject to ongoing review *"by an appropriately experienced and knowledgeable set of personnel"*. On the other hand, Heidi also suggested designers should be involved in some of the early compliance decision-making:

*"Engineers specialise in technology, lawyers specialise in law, there are perhaps technical measures that technical personnel can identify or develop that might satisfy the needs for regulations like GDPR that perhaps lawyers aren't aware of. So in that sense I think technical personnel can assist a legal department in determining what is appropriate for an organisation in terms of protect measures for regulations like GDPR."*

Other potential organisational solutions include investment in training and a vulnerability management team (Grace) or developing tools to support data protection by design (Niaj). In our further discussion below, we will reflect in greater detail on the implications of organisational aspects of cybersecurity.

## Discussion and recommendations

Building on the findings presented in the previous section and also examined against discussions in the literature and the ongoing regulatory initiatives, we now turn to further conceptualisation of the technical, legal and organisational barriers to better cybersecurity management in smart homes. We do this to point towards possible ways forward. It should however be noted that, whilst we present these as distinct technical, legal and organisational considerations, in practice these are not isolated matters but are rather interrelated and overlapping issues. The resolution of the issues in one of those aspects may also likely help address the other two aspects and therefore, tackling all three sets of issues may maximise the regulatory efficacy. For example, an effective, holistic approach to securing devices along the supply chain, as hinted in the ENISA report discussed below, involves endeavours from all those dimensions.

*Technical approaches to security management*

*Role of edge computing for smart home security.* Vendors of IoT products have more power to implement security safeguards at different stages in the IoT supply chain and lifecycle and should not push responsibilities to end-users. Nevertheless, with smart homes we increasingly see responsibility for security being pushed to citizens (particularly so-called domestic data controllers (Urquhart and Chen, 2020) who may have data protection obligations when operating smart devices in the home). This can bring security obligations under Article 32 GDPR (Chen et al., 2020), for example implementing technical and organisational safeguards to mitigate risks to personal data. There is an emerging role for security management tools for the end-user as another key component in supply chain security and managing large scale threats. As such, we are seeing interesting domestic edge-based IoT security management solutions emerging, with commercial offerings and research prototypes (Urquhart and Chen, 2020: 16-17) including *IoT Inspector (2021), Fingbox (2021), CUJO AI (2021), IoT Sentinel (Miettinen et al., 2016), Aretha (Seymour et al., 2020), Homesnitch (OConnor et al., 2019), Sense (F-Secure, 2020), and DADA (Horizon, 2021).*

The approaches vary for these tools but some example approaches include combinations of: analysing and reporting on security of device firmware (*IoT Inspector*); monitoring network activity for security vulnerabilities and then acting by blocking intruders/unknown devices (*Fingbox*) or isolating devices and notifying users (*IoT Sentinel*); classifying device behaviour and network traffic to contextualise what it means to users about data flows (*Homesnitch*); training users in smart firewall management and curating blacklists (*Aretha*); comparing real world IoT device behaviour with manufacturer usage description (MUD) profiles (*DADA*); adapting to emerging threats from new sources (*Sentry*).

*Human-centred security management approaches.* Whilst these technical solutions have appeal by moving security management to the edge of the network in novel ways, more proximate to the end-user, it is worth recalling these need to be situated within the socio-technical context of the home. To do this, we briefly reflect on a couple of studies which highlight risk management, control and usability dimensions of managing domestic IoT security in practice.

Zeng et al. (2017) have observed that users were particularly concerned about the physical security of IoT devices as did Geeng and Roesner (2019) who found users had concerns around safety too, particularly around home co-occupants being able to use IoT resources despite the lack of expertise or account control. This was particularly concerning for users when devices stopped working, e.g. in DIY smart homes.

Zeng et al. (2017: 71) found the levels of technical knowledge users have impacted their awareness of the types of IoT vulnerabilities that exist. For example, skilled users were concerned about encrypted communication channels (HTTPS) but less skilled users focused on weak passwords or unsecured Wi-Fi. They also observed that this level of knowledge shaped how users managed risks from IoT. This ranged from changing behaviour through simple approaches like avoiding speaking in front of the Amazon Alexa smart speaker through to more skilled approaches like creating separate Wi-Fi networks or blocking traffic. Relatedly, Jakobi et al. (2018) note the level of information users seek whilst managing smart homes differs over time. At the beginning there is a desire for more granular information and feedback on current and past behaviour of devices. As years passed, this reduced to wanting information when systems are 'not working, needed their attention, or required active maintenance'.

Geeng and Roesner (2019: 6) explored how interpersonal tensions in a home can impact discussions of control over domestic smart technologies. They found that partners can disagree about third party access via door lock code (e.g., cleaners); roommates can disagree on who controls room temperature via IoT apps; and parents and children can compete for control of an Amazon Echo e.g., with music it plays.

In meeting needs of different users around control, Zeng et al. (2017) argue that systems should 'support multiple distinct user accounts, usability and discoverability of features are critical for secondary, less technical users' e.g., using physical controls and indicators in the home for when being recorded like switching on or off. Similarly, Geeng and Roesner argue that designers need to be sensitive to different relationship types in homes and consider how to make the account creation reflect multiple occupancy with shared devices. They suggest IoT designers need to incorporate 'mechanical switches and controls' for basic device functionality.

These studies help us consider how to reconcile the promise of edge-based security approaches with realities of a domestic IoT deployment context. The practical socio-technical tensions of automating security management in usable ways requires not just to focus on issues like interface design, or longitudinal device management but also more socially fundamental issues around how systems can work for all occupants, not just technically skilled ones and managing impacts on social relations in the home.

*A legal framework supporting co-management of threats*

***Acknowledging smart homes as a cybersecurity frontline.*** Domestic IoT, like many digital services, involves a complex supply chain of vendors, operators, suppliers and third parties. As our participants suggested, where the responsibility for dealing with potential large scale IoT security risks lies needs unpacking. We now consider shifting policies on how responsibility for security considerations is managed across the IoT supply chain and life cycle. This involves examining shifts in critical national infrastructure security regulation through proposed revisions to the EU Network and Information Security (NIS) Directive (2018).

The current NIS Directive focuses on organisations involved in critical national infrastructure with a set of essential and digital services defined within its remit (e.g., cloud computing). It foregrounds the importance of mutual support and collaboration between them to establish and mitigate risks, alongside notification procedures for incidents and implementing safeguards (Urquhart and McAuley, 2018). Emerging reforms in the NIS 2.0 proposals (Commission, 2021b) from December 2020 seek to broaden the scope of the legislation to different bodies, to capture the wider range of actors that need to consider security management (see Article 2 NIS 2.0). It would apply to both essential (Annex 1) and important (Annex 2) entities that are not 'micro or small enterprises'. Certain organisations in the Annexes can be subject to the Directive regardless of size, in a range of circumstances (Article 2(2) NIS 2.0) e.g., where they provide trust, Top Level Domain, or Domain Name Service utilities, are public electronic communications networks or are the sole provider of a service in the member state.

Whilst the list of essential providers mirror those in the current NIS legislation, including transport, energy, banking, financial markets, health, and water distribution, the list of digital infrastructure services would cover broader. It would include data centres, content delivery, DNS and TLD name registries, exchange points, providers of public electronic communications networks. Interestingly, in addition to providers of online marketplaces, online search engines and social networking platforms, the manufacturers of computer, electronic, electrical and optical products would now be included as *important* entities also subject to NIS obligations under the proposal.[5] In part, this recognises growing security obligations that need attention as industry 4.0 and smart manufacturing emerge, but it also shows the complexity of ensuring security across supply chains. Even since 2016 (when the original NIS Directive was passed), the entities which have responsibilities for critical national infrastructure needs to be more broadly construed.

It also highlights the importance of managing security across the supply chain. For example, as part of developing EU cybersecurity certification schemes (Article 21 NIS 2.0) for products, services and processes to demonstrate compliance with requirements for technical and organisational measures being deployed within an organisation, it includes a focus on supply chain security as one measure (Article 18). NIS 2.0 proposes that this includes "*security-related aspects concerning the relationships between each entity and its suppliers or service providers such as providers of data storage and processing services or managed security services.*" (Article 18(2)(d)) It also states there need to be EU coordinated security risk assessments involving ENISA and the Commission for critical supply chains. (Article 19)

It also brings some measures which will help both establish and manage responsibilities to minimise scope for large scale infrastructural vulnerabilities. This includes new *vulnerability disclosure* approaches. As the proposed legislation states in Article 6 (1), Member states will coordinate a vulnerability disclosure scheme, where the Member State CSIRT[6] operate this, cooperating with other CSIRTs in relation to notified vulnerabilities. This is in conjunction with an ENISA-run EU wide registry which includes: "*information describing the vulnerability, the affected ICT product or ICT services and the severity of the vulnerability in terms of the circumstances under which it may be exploited, the availability of related patches and, in the absence of available patches, guidance addressed to users of vulnerable products and services as to how the risks resulting from disclosed vulnerabilities may be mitigated.*" (Article 6(2)) Given the lucrative trade in zero-day vulnerabilities and stockpiling of these by security agencies, cybercriminals and nation states, this disclosure strategy may have significant gaps in coverage but at least it showcases a new approach to coordinating response to vulnerabilities (e.g., beyond ethical vulnerability disclosure processes) (Maurushat, 2013). This is key, where vendors play a role in reporting vulnerabilities so these can be patched and risks of large-scale impacts can be avoided.

Concurrently to the NIS 2.0 proposals, ENISA (2020) has been doing some work establishing new approaches to thinking about security across IoT supply chains. Their recent report from December 2020 highlights several priorities for ensuring responsibilities are well managed. This looks across the IoT supply chain, from conceptual to development to production to utilisation, support and retirement stages. For the purposes of this discussion, we highlight 3 elements that pertain to addressing critical infrastructural security risks, particularly in the latter stages once IoT is deployed in use.

---

[5] See section C division 26 of NACE Rev 2 for the full long list (computer, electronic and optical) and Division 27 (electrical equipment) (https://ec.europa.eu/eurostat/documents/3859598/5902521/KS-RA-07-015-EN.PDF). This also includes other manufacturers of transport equipment, machinery etc.

[6] Computer Security Incident Response Team (e.g., in UK, the National CSIRT is the National Cybersecurity Centre, but there are numerous other private sector and governmental CSIRTs. For a full list, see https://www.enisa.europa.eu/topics/csirts-in-europe/csirt-inventory/certs-by-country-interactive-map#country=United%20Kingdom )

Firstly, with '*end user operation and service provision*', ENISA highlight the importance of elements such as:

- provision of management and technical support resources to ensure security across the IoT device lifecycle;
- ensuring compromises in user experience and convenience are balanced against security approaches;
- ensuring that IoT operators have training to guard against risks from misuse or misconfiguration;
- use of access controls; seeking to increase uptake of security measures.

Secondly, in *ongoing technical support and maintenance*, they primarily highlight the importance of providing over the air, secure maintenance tools and patching. This is another key area of concern in the UK proposals, as mentioned above, particularly vendors specifying for how long they will provide update support. This is related to the third point around data removal for end of life with devices. Whilst many of these provisions are drafted with organisations and business users in mind,[7] as we discuss below, domestic operators of IoT devices increasingly have obligations around data processing, and thus these strategies demonstrate the importance of not pushing all responsibilities just to end-users, but that ongoing security support from vendors is critical. The nature of support vendors of IoT products and services need to provide is not well defined, but the ENISA guidance shows possible directions for this, to mitigate risks posed by domestic IoT being managed purely by end users (as part of the IoT supply chain).

***Certification and labelling mechanisms.*** A common theme picked up by the respondents was the difficulty in complying with inconsistent standards issued by different bodies and applied in different countries. Fully resolving this issue would ultimately entail harmonisation of different standards, but in practice, sector- and country-specific efforts can help mitigate this issue for IoT vendors, especially SMEs who do not necessarily have the resources to cope with multiple standards. This is where certification schemes may come into play, as suggested by some interviewees and also in work by Piasecki et al. (2019). A certification body approved by the competent regulator – which can be a sector-specific one – may compile the applicable legal requirements and cybersecurity standards, and turned them into a set of practical, demonstrable and verifiable guidelines tailored to the sector concerned.

There are international certification schemes already in place designed to evaluate the security level of ICT systems. One of the most prominent schemes is Common Criteria (CC, also known as ISO 14508), which built largely on the European standard ITSEC developed by the UK, Germany, France and the Netherlands in 1991 (Houmb et al., 2010). There have been suggestions (Nitu, 2019), and indeed practices (Kang and Kim, 2017), in certifying IoT devices with CC, but the number of consumer products on the list of CC-certified products remains minimal. CC is also subject to criticisms as to its suitability for the IoT sector. Baldini et al.

---

[7] ENISA Report states its scope on p7 in s1.2 as "all the stages of the IoT supply chain, defined as a holistic system of organizations, people, technology, processes, information, and other physical and virtual resources involved in the whole lifespan of any IoT product or service, from the conception to the end customer supply and the end of the product life cycle."

(2016), for example, point out that the time and costs involved in the process to acquire a CC certificate are hard to justify in the fast-developing IoT market. Also, the chosen CC protection profile does not always match the needs of the user, and may even conflict with other devices put into the same system (Baldini et al., 2016). More importantly, Matheu et al. (2019) highlight the challenge that version-specific schemes, such as CC, would be incompatible with the common industrial practice of updating or patching IoT products on a regular basis.

In the UK, the National Cyber Security Centre (NCSC) has stopped issuing CC certificates since October 2019 (NCSC, 2019). The NCSC offers alternative certification services (NCSC, 2021b), but the only one targeting consumer products is Commercial Product Assurance (CPA), which only covers smart metering products (NCSC, 2021a). In this regard, there might be scope for a more targeted, government- or industry-led security certification schemes designed for domestic IoT products. Given the significant overlap between cybersecurity and data protection standards, there is also a possibility of alignment, or even integration, between IoT security certification schemes with data protection certification schemes. The GDPR sets out a voluntary certification scheme and lays down the conditions that the certificates, the certification bodies and the accrediting authorities must meet. (Articles 42, 43) This approach is viewed as a "monitored self-regulation" model by Lachaud (2018), who also calls for further improvements of the current regime as dictated by the GDPR. With specific regard to IoT products in relation to both security and data protection matters, an effective certification scheme must address the supply chain issue highlighted above, examining whether consumers (in a B2C context) or IoT vendors (B2B) would be the more appropriate audience of such certificates.

A more consumer-facing and arguably lighter-touched alternative to certification is labelling, sometimes known as "self-certification", as preferred in the UK government's initial regulatory proposals (DCMS, 2019), albeit entirely turned down at a later stage (DCMS, 2020c). It should however be noted that there are significant differences between certification and labelling mechanisms: For one thing, the labelling scheme proposed by the government is meant to be a mandatory rather than voluntary regime, meaning that all IoT products must bear a security label with all the details mandated by law. For another thing, there is no designated certification body under the labelling scheme, with the devices self-certified by the manufacturers and retailers.

The shift of the government's preferred regulatory option away from the labelling scheme perhaps resulted from the "diverse range of opinions" received during the consultation (DCMS, 2020b). Such a disagreement is also evident among our interviewed experts, as reported above. Partially echoing the food safety regulation analogy came up with by the interviewees, there are lessons that the design of a labelling scheme can learn from the regulatory experience in, for instance, mandatory disclosure of food nutrition information. Indeed, work by Kelley et al. (2009) and Cranor (2012) speaks of a "privacy nutrition label", which draws on food nutrition labelling regimes with a view to informing the design of similarly standardised labels or icons. It has been empirically proved that privacy notices for IoT products are poorly accessible, making it hard for consumers to make meaningful use of them (Paul et al., 2018; Perez et al., 2018; Emami-Naeini et al., 2019). As such, making information available to consumers of IoT products in a truly legible and usable manner should be a priority for policymakers. Another lesson that can be learned from the food industry comes from the fact that unsafe and unhealthy food are regulated differently: The former is prohibited whereas the latter is warned against through proper disclosure of information. In a similar vein, the UK government's first two regulatory options – "mandating a security label" and "mandating the top 3 guidelines" – are not necessarily incompatible with one another. There is scope for prohibiting consumer IoT products not compliant with the top 3 guidelines while at the same time requiring vendors to declare the level of compliance with the rest of the guidelines through a standardised format. Further research, however, will be needed on how this can be implemented in practice.

### *The need for organisational strategies*

***Competing priorities in organisations.*** Driven by commercial interests, maximising profits is perhaps unsurprisingly a common objective for most IoT vendors. At the same time, ensuring products are secure and supporting users to manage security can also form part of the organisations' strategic priorities. These two goals are not always considered fully compatible with each other by all organisations, many of whom are most likely to prioritise profit over security when they see them as in conflict. How much an organisation would have the incentive to invest in security would therefore depend on the extent to which efforts in improving security can be aligned with the financial sustainability of the organisation.

It has been pointed out by multiple interviewees (Alice, Evan and Mallory) that for SMEs or start-ups producing low-cost connected gadget, it tends to be harder to find the scope to internalise the costs of developing secure products as well as maintaining them for the longer term. Surviving in a relatively competitive market by having their products rolled out as quickly and inexpensively as possible is regarded as the top priority of many of the smaller organisations. At the same time, most interviewees also think there should not be such a misalignment. In the long-term, enhancing IoT product security can be in line with the commercial interest of the organisation. The absence of appropriate security planning early on in the development process can cause great costs to organisations, including costs to fix or even recall insecure products, reputational and trust damage among consumers and within the supply chain, and in serious cases, direct financial losses due to consumer claims or fines imposed by regulators.

Such competing priorities also manifest in how different departments communicate and work together within the organisation. As highlighted in our findings above, many interviewees reported observations of compliance being an after-thought and often dealt with solely by the legal department. With security and data protection largely seen as compliance requirements to be dealt with mainly by the legal department, product development often does not involve the legal department early on. The development department may have some understanding of what is required by law or technical standards, their performance is measured mainly by the feasibility, marketability and profitability of the product design. Even if the legal department is involved, the language barriers between the two departments sometimes prove to be too hard to overcome. Meaningful engagement with a view to enhancing compliance with product security requirements across the organisation remains a significant challenge.

In this regard, a change of corporate culture in how the security of products should be treated as a priority, as suggested by interviewees, would make a difference. To make this change happen, strong leadership in the executive management would be critical. As Heidi pointed out, the conflict of priorities between departments is "more a culture thing and more a management style". Promoting effective inter-department communications and collaborations across the IoT sector is thus not only a matter of improving organisational governance internally, but can significantly heighten the overall level of security of IoT products, which in turn improves cybersecurity for individual consumers as well as the wider network.

***Organisational structure matters.*** One point discussed with regard to the regulatory landscape above was the different connotations of the "by design" terminology in adjacent regulatory spaces. While the concept of "data protection by design" upheld in Article 25 of the GDPR emphasises both organisational and technical measures, the "secure by design" approach in IoT cybersecurity focuses solely on the technical aspect. In the context of regulating AI – something comparable to IoT with both being data-driven technologies – there have already been discussions on how management structure of firms developing AI solutions can impact the effectiveness of accountable AI policies. The EU's White Paper on AI, for example, proposes "setting up a broad-based public private partnership, and securing the commitment of the top management of companies" (Commission, 2020: 7). The UK ICO (2020: 13)'s draft AI Auditing Framework also has a strong focus on the governance structure of organisations, explicitly suggesting that important decisions regarding AI risk management cannot be delegated to data scientists or engineering teams but should be address directly by the senior management. More specifically, it was suggested that:

> *"To do so, in addition to their own upskilling, they need diverse, well-resourced, teams to support them in discharging their responsibilities. You also need to align your internal structures, roles and responsibilities maps, training requirements, policies and incentives to your overall AI governance and risk management strategy."* (ICO, 2020: 13)

In contrast, the Secure by Design initiative has largely overlooked these important organisational dimensions in the recommended practices. None of the 13 guidelines in the Code of Practice for Consumer IoT Security addresses how to improve corporate structure to support implementation of the technical requirements. The same lack of coverage on organisational considerations is seen also in the ETSI Technical Specification on Cyber Security for Consumer Internet of Things. Only more recently, as discussed above, ENISA (2019: 51-52) addressed the issues surrounding the supply chain and recommended a range of good practices for security of IoT, which does cover the "people" aspect, including suggestions addressing "training and awareness", "roles and privileges" and "security culture". These recommendations, however, are yet to be translated into the UK's regulatory practices.

Such a renewed understanding of "secure by design" should be translated into policymaking. Rather than viewing organisations as black boxes, policymakers should begin thinking about how to unpack the internal workings of organisations developing and operating IoT products and services, and how engagements, recommendations and even interventions could facilitate organisations to secure their smart products, and to support consumers to secure their smart homes. This should also form part of the wider paradigmatic shift to the human-centred security approach to regulating consumer IoT product, as highlighted by Piasecki et al. (2019).

## Conclusion

Our study with 13 experts from the industry and governmental bodies sheds light on the socio-technical challenges facing the consumer IoT sector, specifically focusing on the technical, legal and organisation aspects. The findings show that IoT vendors, especially SMEs, experience major obstacles on all those three dimensions, leading to further reflection on what is lacking in the current regulatory environment and what is needed to support and incentivise manufacturers and retailers to not just make their consumer IoT products more secure, but also to make the security of their products genuinely manageable by average end-users. Some of the challenges are interrelated, which can make the coping strategies more complex, but at the same time, could also mean that a holistic approach may be possible and necessary. The ongoing regulatory efforts in the UK are overlooking some of those socio-technical factors, which could potentially undermine the efficacy of such initiatives. We believe there is need to prioritise the following approaches to redress this, including: adopting a human-centred approach to security across the IoT supply chain from organisations to homes; ensuring vendors develop technical and organisational support structures that attend to the growing role of domestic data controllers in managing smart home security; expanding the scope of cybersecurity laws to forecast and manage risks across the lifecycle of IoT devices, alongside improving capacity to responsively collaborate to tackle emerging IoT risks; rolling out certification and labelling schemes that increase consumer trust in devices; and promoting security-minded organisational structures. We believe policymakers should give due regard to these elements in order to better secure the smart home.